# Feedback Loops in Open Data Ecosystems

**D. Rudmark**
RISE Research Institutes of Sweden

**M. Andersson**
RISE Research Institutes of Sweden

*Abstract— Public agencies are increasingly publishing open data to increase transparency and fuel data-driven innovation. For these organizations, maintaining sufficient data quality is key to continuous re-use but also heavily dependent on feedback loops being initiated between data publishers and users. This paper reports from a longitudinal engagement with Scandinavian transportation agencies, where such feedback loops have been successfully established. Based on these experiences, we propose four distinct types of data feedback loops in which both data publishers and re-users play critical roles.*

OPEN DATA has, since its inception, held a promise to both create wealth and strengthen democratic institutions. One of these opportunities is to let external private actors refine the data into end-user services and, by extension, societal benefits. A core, enabling characteristic of open data is being an inherently *open* system, where few natural opportunities for user feedback exist. Since many of the early promises of open data have yet to materialize, this lack of feedback loops has been pointed out as one crucial reason[1]. At the same time, there are open data ecosystems that seem to have resolved these issues. For example, open datasets are currently re-used by global platform leaders like Google and Apple as crucial input in their data-driven service design, used by hundreds of millions of end-users in applications like vehicle navigation systems and multi-modal travel planners. In this paper, we briefly examine how open data quality issues are being tackled among transportation agencies and report a set of strategies these agencies have employed to manage feedback loops to the data curating process.

More than 20 years ago, Orr [2] argued that the core mechanism to ensure data quality is to improve the *use* of data. In open data, such use-oriented quality properties include data currency, understandability, and accuracy[3]. For organizations holding datasets that are subject to open data publishing, this means moving from a simplistic assumption of merely publishing data to regularly engage with users of that data to find out what they are actually doing with it. Whether the purpose is to enhance democratic institutions [1, 4] or improving systems [2], establishing a well-functioning feedback loop building on *data use* is a key part of successful open data quality management.

OPEN DATA FEEDBACK LOOPS

Following our longitudinal engagement with Scandinavian transport agencies and their work with two datasets – static public transport data and dynamic traffic information, we found that these organizations indeed had succeeded in establishing robust and quality-assuring feedback loops. While they differed significantly concerning the involved actors and level of formality, we found, as suggested by Orr [2], that two core components enabled these feedback loops. First, recurrent and dependable use of the open data was put in place, typically through high-usage end-user applications. Second, these feedback loops depended on having an actor taking on the responsibility to initiate the feedback loop when data quality was insufficient.

What was more surprising was that these components varied depending on who took on what role. We found that the recurrent *use context* could be managed by either (as expected) external actors – but also by internal data publishers. Moreover, we found that the *feedback loop initiator* could either be an external re-user or the open data publisher themselves. In what follows, we present four cases in which specific combinations can be discerned (Figure 1).

Box 1. Background studies

|                          |                  | Use context        |                                          |
|--------------------------|------------------|--------------------|------------------------------------------|
| Feedback loop initiator  | Data Publisher   | *Data Dogfooding*  | *Third-Party Application Monitoring*     |
|                          | Data Re-user     | *Community Curation* | *External Quality Proxy*               |

Figure 1. Open Data Feedback Loops

### Data Dogfooding

We identified the first type of feedback loop within the Swedish Transport Administration. This governmental organization has the overall responsibility for both the physical and digital transport infrastructure in Sweden. As such, the administration is responsible for communicating traveler-critical information like passage times, departure platforms, and potential delays for passenger trains; traffic-disturbing road accidents and road works; and timetables and real-time updates of road ferries. Given how critical this information is for travelers, there is substantial external demand for this information. Accordingly, the STA has since 2014 published this information as open data to accommodate external service development better and thereby meet traveler demand.

However, while external demand and use of this information indeed are high, the user invoking the most API calls is the STA themselves[1]. This usage pattern is because the STA has chosen to use the exact same API for their public-facing services as they offer to any open data user. We refer to this type of feedback loop as *data dogfooding* [5]. By dogfooding, the STA addresses several quality challenges surfacing for open data publishers once data is released to the public. First, as the open API serves as the basis for several mission-critical STA applications, dogfooding installs internal incentives to maintain non-functional quality aspects of the API, such as latency, up-time, and consistency. Second, any updates to the APIs functionality must be internally verified as the STAs applications are contingent on consuming these upgraded interfaces.

### Third-Party Application Monitoring

This feedback loop involves the STA and its traffic data and several navigation service companies using this data in their end-user apps. Keeping road users informed about traffic is a core task for the STA. While they operate large traffic management centers in large urban areas, in recent years, navigation services, such as TomTom, Google Waze, Inrix, Here, and others outside of STA have grown indispensable for road users. This resource can be accesses either through the comprehensive XML-based DATEX II standard interface or a leaner API to get data about planned and unplanned traffic disturbances. However, on occasion, the STA could observe that events were not represented as they would have expected in these external navigation apps. The STA realized that securing quality internally did not seem to be enough. They needed to know how their data was being used in situ by the navigation services. At the same time, navigation service providers could not be expected to provide detailed data on deviances. They had not reacted to these faulty results thus far.

Taking a data-in-use perspective, the STA initiated their own investigation to catalog deviances in navigation services and categorizing them. It soon became apparent that it was difficult to show a specific pattern involving certain actors. Instead, the errors seemed more or less randomly distributed among the services. Furthermore, after initiating dialogue with service providers using the documented deviations as a base, it was found that events in the STA API were often geo-referenced by a single point rather than a segment of the road. Consequently, this dataset design made it difficult for services to interpret the data. This *third-party application monitoring* feedback loop then instigated an improvement project to alter geo-referencing practices within the STA.

---

[1] In August 2020 approx. 147 000 000 API calls was handled by the STAs open API. Out of these, approx. 41 000 000 calls originated from STAs applications, and the remaining 106 000 000 calls were invoked by externally developed services.

### Community Curation

Our third example was found within Entur, a state-owned organization that collects, transforms, and publishes open data for all of Norway's public transport and concerned geospatial data. Geospatial open data on road networks is a valuable asset that can optimize public transport trips as travelers need help to calculate the best trip options within the public transport network and connecting walking segments to and between stops and stations. Entur relies heavily on the open, collaborative map OpenStreetMap (OSM) to calculate such walking segments. However, while OSM's network quality in urban areas is typically both detailed and current, quality in rural areas is often lacking. To this end, several actors, including Entur, are importing various open data into OSM.

Since the governance processes of OSM only includes curation of the data after changes are uploaded, any perceived deviations from reality unfold as a joint curation among active contributors within OSM. For example, such deviations can be faulty interconnections between road segments (either missing when there are connections or a connection in OSM while the segments in practice are separated by, e.g., a fence). Other examples include contributions with specific purposes (e.g., a landowner that does not want traffic on its road and tags them as undrivable). In rare cases, deliberate sabotage edits also occur. When this happens, the community engages in an ongoing dialogue that we refer to as *community curation*. These discussions on what exists, in reality vs. the map, are initiated by experts with local knowledge of current conditions, notifying that existing data may be faulty. To draw on the ecosystem of actors within OSM, Entur has been very active in the Norwegian OpenStreetMap community, having one employee spending significant time on the data import and interactions with community experts. Among other benefits, this engagement has created an opportunity for Entur to correct outdated data points[2].

### External Quality Proxy

Our final feedback loop was identified between Samtrafiken[3] and Google. A common standard for publishing transit data is GTFS [6], that allows public transport organizations[4] to export their timetables, stop infrastructure,

---

[2] One such example can be found here: https://www.openstreetmap.org/note/2118234

[3] Samtrafiken collects, transforms, and publishes open data for Sweden's entire public transport operations

[4] Currently, the open public transport data register OpenMobilityData https://transitfeeds.com/feeds lists approx. 850 public transport agencies publishing data in the standard GTFS format

and route networks into various routing services[5].

However, as owners of static transit data export legacy system databases into GTFS feeds, there is a risk that errors are introduced into the feeds. Moreover, these errors may elude public transport data publishers since they often are not users themselves of this data. In the GTFS case, we thus noticed how transit agencies instead relied on Google Inc. as the vetting body of GTFS feeds, and we refer to this type of feedback loop as *external quality proxy*. In practice, this proxy role is exercised by Google through requirements that all GTFS feeds are subjected to automated consistency checks before the feed is allowed to be imported into Google Maps. In practice, this means that the feed import is stalled until the agency addresses the errors and the feed complies with Google's verification. Since Google Maps is a critical end-user application for many public transport actors, such errors are thus typically fixed quickly.

As GTFS feeds are often published as open data, all re-users can enjoy the quality standards required by Google. According to the Open Data Manager at Samtrafiken, using Google as external quality proxy has led to significant quality improvements on Samtrafikens GTFS feed and that more upstream, and previously undetected, data errors have been detected.

CONCLUDING RECOMMENDATIONS

As our investigation shows, there are rife opportunities for data publishers to establish feedback loops based on actual use at different locations in the value chain and involving various actors. In what follows we present four such opportunities based on our research:

*Use your own open data.* Owners of open data are typically the most knowledgeable about what aspects that should be present in high-quality datasets. Consequently, avoiding making open datasets a new, separate publication targeting only external users has proven a requisite quality measure in several of our studied public transport organizations. For instance, when Stockholm Public Transport started to use the open data API in 2012 for their external services, SL identified quality issues in the data itself. Despite the API being operational for more than a year and serving hundreds of thousands of end-users of externally developed services, data around passing stops on a public transport journey was sometimes missing.

*Investigate how your data is re-used.* Another action following our research is investigating how re-users in practice interpret and present open data to their users. In our case data, this type of action was particularly useful for multinational services building on standardized open data. Here, it appeared as service providers lacked resources for extensive cross-checking of dataset imports. To illustrate with an example in this vein, consider the parking app Parkopedia that presents parking options for its users. According to its web page, it currently uses data from 15000 cities [7]. As such, we speculate that Parkopedia may have difficulties to rigorously verify the data quality from all these cities. If data owners probed the presentation of data from their particular city, quality deficits might emerge.

*Cooperate with the community.* While data owners drive the previous two recommendations, we found that more reciprocal quality-enhancing feedback loops can be established through cooperation with civic enthusiasts. This has been especially useful for geospatial data where the "many eyeballs of the community" have corrected map deviations from the lived environment. For instance, as Denmark released its address data openly, authorities collaborated with the OSM community that quickly discovered and fixed faulty data points in the open dataset[8]. Similarly, when Zürich released its parking spaces as open data, citizens swiftly reacted to data inaccuracies [9].

---

[5] These include apps like Google Maps, transit.app, and Moovit.

*Find an external expert to assess your data*. Finally, sometimes feedback loops can be established when a trusted authority can validate the quality of open data. For instance, in the transportation domain, the navigator app Waze has a data-sharing program with cities[6]. Here, cities participating in this program may push road closures, incidents, and snow plowing vehicle positions to Waze for presentation to the app's users. Waze then validates the data and feedbacks quality deficiencies to the road authority as part of the publishing process. Once corrected, the quality-enhanced data can also be published for open re-use.

# ■ REFERENCES

[1] M. Janssen, Y. Charalabidis, and A. Zuiderwijk, "Benefits, adoption barriers and myths of open data and open government," *Information systems management,* vol. 29, no. 4, pp. 258-268, 2012.
[2] K. Orr, "Data quality and systems theory," *Communications of the ACM,* vol. 41, no. 2, pp. 66-71, 1998.
[3] A. Vetrò, L. Canova, M. Torchiano, C. O. Minotas, R. Iemma, and F. Morando, "Open data quality measurement framework: Definition and application to Open Government Data," *Government Information Quarterly,* vol. 33, no. 2, pp. 325-337, 2016.
[4] J. Attard, F. Orlandi, S. Scerri, and S. Auer, "A systematic review of open government data initiatives," *Government Information Quarterly,* vol. 32, no. 4, pp. 399-418, 2015.
[5] W. Harrison, "Eating Your Own Dog Food," *IEEE Software,* vol. 23, no. 3, pp. 5-7, 2006.
[6] Google Inc. "General Transit Feed Specification Reference," 2021-01-24; https://developers.google.com/transit/gtfs/reference/.
[7] W. Trinkwon. "Parkopedia indoor mapping prevents sat-nav 'blackouts'," 2021-07-02; https://www.autocar.co.uk/car-news/industry-news-technology/parkopedia-indoor-mapping-prevents-sat-nav-%E2%80%98blackouts%E2%80%99.
[8] J. McMurren, S. Verhulst, and A. Young. "Denmark's Open Address Data Set - Consolidating and Freeing-Up Address Data "; https://odimpact.org/case-denmarks-open-address-data-set.html.
[9] R. Strauman. "Win-win: Schliessen des Open Data-Feedback-Loops," 2021-07-05; https://digital.ebp.ch/2016/11/15/win-win-schliessen-des-open-data-feedback-loops/.

**Daniel Rudmark** works as senior researcher at RISE Research Institutes of Sweden on issues relating to open transport data publication and re-use. He has worked with several actors within the Swedish public transport sector and internationally in the Nordics, Rio de Janeiro, Dar es Salaam, and Mysore. Besides doing applied research at RISE, he has consulted for the World Bank on open transport data and completed a Ph.D. focused on open data platforms.

**Magnus Andersson** works as senior researcher at RISE Research Institutes of Sweden, focusing on digital innovation and data sharing within the automotive and transportation sectors. He has specialized in heterogeneous digital ecosystems and has accumulated experiences doing applied action-oriented research together with global firms and government agencies for the past two decades.

---

[6] See https://www.waze.com/wazeforcities/ for details